# COMMENTS ON "SUB-$k_\mathrm{B}T$ MICRO-ELECTROMECHANICAL IRREVERSIBLE LOGIC GATE"

LASZLO B. KISH

[1] *Department of Electrical and Computer Engineering, Texas A&M University, College Station, TX 77843-3128, USA*

*laszlo.kish@ece.tamu.edu*

**Abstract.** In a recent article, *Nature Communications* **7** (2016) 12068, the authors claimed that they demonstrated sub-$k_\mathrm{B}T$ energy dissipation at elementary logic operations. However, the argumentation is invalid because it neglects the dominant source of energy dissipation, namely, the charging energy of the capacitance of the input electrode, which totally dissipates during the full (0-1-0) cycle of logic values. The neglected dissipation phenomenon is identical with the mechanism that leads to the lower physical limit of dissipation (70-100 $k_\mathrm{B}T$) in today's microprocessors (CMOS logic) and in any other system with thermally activated errors thus the same limit holds for the new scheme, too.

*Keywords:* Energy dissipation of control; Dissipation limits in logic gates; Error probability.

In a recent article [1], the authors claim that they demonstrate sub-$k_\mathrm{B}T$ energy dissipation at elementary logic operations. In the experiments, they use Micro-Electromechanical Cantilevers (MEC) where the attractive electrostatic forces toward the input electrode(s) control the position of the cantilever tip.

The Authors discuss an OR gate however, for the sake of simplicity but without the loss of generality, we investigate a Follower logic gate in this paper. Two separate tip positions define the two logic values, 0 and 1, respectively. These logic values and tip positions correspond to driving voltages 0 and $U_1$ on the input electrode. The energy dissipation (due to friction) in the cantilever is measured during changing the logic value and the Authors correctly conclude that the energy dissipation due to friction can be made smaller than $k_\mathrm{B}T$.

This is the point where we must ask:

*Does this approach account for all the major energy loss phenomena determining the lower limit of energy dissipation or is there any dominant but neglected component?*

The answer is straightforward:

The considerations in [1] neglected the energy dissipation due to charging and discharging the input capacitance, which are the same dissipation phenomena [2-6] that determine the lower limits of dissipation (70-100 $k_\mathrm{B}T$) [3] not only in today's microprocessors (CMOS logic) but also in any other system with thermally activated errors [2-6]. Thus the same dissipation limit (70-100 $k_\mathrm{B}T$) holds for the new scheme in [1], too.

*No sub-kT logic operations, except erasure.*

While the flaw in [1] is now obvious due to the above arguments and no further considerations are needed to deny the validity of the sub-$k_B T$ energy dissipation claim, for the Readers who are less familiar with the topic, we provide more details below. In voltage-controlled logic, see Figure 1, when the switch $S_1$ is closed, the voltage on the capacitor is changing from the value 0 to $U_1$ while the logic (bit) value is switching between 0 and 1.

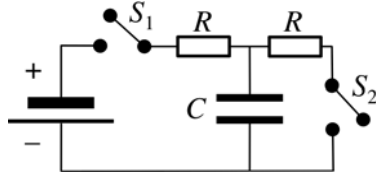

Fig. 1. Elementary circuit model to show the major energy dissipation in voltage-controlled logic [3].

Thus, in the case of the 0 ==> 1 bit value change, the energy of the capacitor is changing from zero value to

$$E_1 = 0.5 C U_1^2 \quad , \tag{1}$$

while exactly the same amount of energy ($E_1$) is dissipated in the resistor $R$ due to its heating by the charging current. Here $C$ represents the input capacitance (the gate capacitance in CMOS logic) and $R$ is the resistance of the closed switch (the source-drain resistance in CMOS). During erasure by resetting, $S_1$ is off, $S_2$ is closed, and again energy $E_1$ is dissipated in the other resistor. However, the resistors create Johnson noise (thermal noise) leading to thermally activated bit errors depending on the magnitude of $U_1$, which separates the logic states 0 and 1. Considerations based on these effects result in the following formula for the lower limit of energy dissipation in arbitrary systems with thermally activated errors [2-6]:

$$Q_\tau \approx k_B T \ln \frac{1}{\varepsilon} \quad , \tag{2}$$

where $\varepsilon$ (<0.5) is the error probability during the observation time, and the formula is valid in the short observation time limit, that is, when the observation time of error events is less than the correlation time $\tau$ of thermal fluctuations activating the errors (the Johnson noise on the capacitor). The value of $Q_\tau$ is around 70 $k_B T$ at today's error probability expectations [3].

In the long observation time, $t_o$, limit, Equation 2 is slightly modified [5,6]:

$$Q_t \approx k_B T \left[ \ln\left(\frac{1}{\varepsilon}\right) + \ln\left(\frac{t_o}{\tau}\right) \right] \quad , \tag{3}$$

which results in a slightly higher energy dissipation, $Q_t \approx 100 k_B T$ with today's typical bandwidth and $t_o = 1$ year .

The logic gate described in [1] is also a voltage-controlled gate with the unavoidable Johnson noise during charging and discharging the input capacitance [6], thus, Equations 2,3 and the corresponding dissipation of 70 - 100 $k_B T$ holds there, too.

However, the argumentation is invalid because it neglects the dominant source of energy dissipation, the 50% loss of the charging energy of the capacitance of the control gate and the total loss of the remaining part during discharging. The neglected dissipation phenomenon is identical with the mechanism that leads to the ultimate dissipation limit in today's microprocessors (CMOS logic), thus the same limit holds for both.

To preempt the question if there is a way to charge and discharge a capacitor without the loss of energy shown above we show the resonant circuit solution used in switching power supplies, see Figure 2. Suppose $C_1$ is charged, $C_2$ is not. Closing $S_1$ causes a sinusoidal current flow in the coil $L$. At the peak value of the current, all the energy is magnetic and $C_1$ has discharged state. Then $S_2$ closes and $S_1$ opens. When the current reaches the zero value, all the energy is transferred into $C_2$. However, such energy saving scheme is efficient only at large (>>70 $k_B T$) charging energies. The situation is much worse at the logic gate energy range because then two new switches must also operate, each one with energy dissipation similar to the energy we want to save!

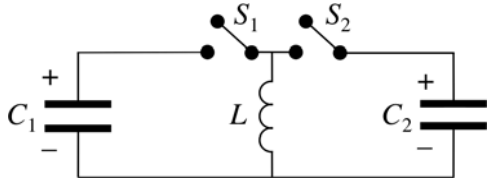

Fig. 2. "Tank" LC circuit example to bounce the charging energy in a lossless way between two capacitors with timed switches and LC resonator. Due to the energy dissipation of controlling the switches, it fails at the charging energy range where logic gates operate. All similar efforts require to increase the number of switching events thus lead to increased energy dissipation at the energy levels of logic gates.

Finally, without arguments or details, we summarize our general results [7,8] in the related matter of energy dissipation in memories and logic gates:

*i. Information entropy cannot be interrelated with dissipation, in general* [8].

*ii. Writing of data (running of logic gates) is dissipation costly.*

*iii. Erasure, on the other hand, can be completely dissipation-free, in special cases* [7] .

There are interesting alternative considerations with related conclusions, see, e.g. [9,10].

**References**

*No sub-kT logic operations, except erasure.*